\documentclass[12pt]{article}
\usepackage{amsfonts}
\usepackage{latexsym,amsmath}
\usepackage{amssymb,array}
\usepackage {hyperref}
\usepackage{color}
\makeatletter \oddsidemargin 0in \evensidemargin 0in \textwidth 16cm 
\RequirePackage[dvips]{graphicx} \textheight 20cm
\begin{document}
\newcommand{\ga}{\gamma}
\newcommand{\la}{\lambda}
\newcommand{\wt}{\widetilde}
\newcommand{\ov}{\overline}
\newcommand{\La}{\Lambda}
\newcommand{\ri}{\Rightarrow}
\newcommand{\bv}{\vee}
\newcommand{\bi}{\wedge}
\newcommand{\e}{\exp}
\newcommand{\eq}{\Leftrightarrow}
\newtheorem{thm}{Theorem}[section]
\newtheorem{coro}{Corollary}[section]
\newtheorem{remark}{Remark}[section]
\newtheorem{counterexample}{Counterexample}[section]
\newtheorem{note}{Note}[section]
\newtheorem{definition}{Definition}[section]
\newtheorem{example}{Example}[section]
\newtheorem{lem}{Lemma}[section]
\newtheorem{pro}{Proposition}[section]
\renewcommand{\theequation}{\thesection.\arabic{equation}}
\title{On Some Generalized Orderings: \\In the Spirit of Relative Ageing}
 \author{Nil Kamal Hazra and \;Asok K. Nanda\footnote{e-mail: asok.k.nanda@gmail.com, asok@iiserkol.ac.in, corresponding author.}
 \\Department of Mathematics and Statistics
 \\IISER Kolkata, Mohanpur Campus
 \\Mohanpur 741252, India}
 \date{October, 2014}
\maketitle
\begin{abstract}
 \hspace*{0.1 in}We introduce some new generalized stochastic orderings (in the spirit of relative ageing) which compare probability distributions with the exponential distribution.
 These orderings are useful to understand the phenomenon of positive ageing classes and also helpful to guide the practitioners when there are crossing hazard rates and/or crossing
 mean residual lives. We study some characterizations of these orderings.
 Inter-relations among these orderings have also been discussed.
\end{abstract}
\section {Introduction}
\hspace*{0.14 in}
Lehmann~\cite{le6} proposed proportional hazard (PH) rate model (commonly known as Cox's PH model, see Cox~\cite{cox}) 
which is very useful to analyze the failure time data in reliability and survival analysis. 
Later, Zahedi~\cite{za6} introduced proportional mean residual life model which is a parallel concept to Cox's PH model.
 In many real life situations, the comparison of two crossing hazard rates and/or crossing mean residual lives has been observed, 
 see, for instance, Pocock {\it et al.}~\cite{pgk6}, Champlin \emph{et al.}~\cite{cmeg}, Begg \emph{et al.}~\cite{bmbco}, Mantel and Stablein~\cite{ms6}, Gupta and Gupta~\cite{gg6},
 and Bekker and Mi~\cite{bm6}.
Some methods under the Cox proportional hazards framework have been developed to deal with the crossing hazard rates
problem (cf. Liu \emph{et al.}~\cite{lqs6}). Sengupta and Deshpande~\cite{sd2} discussed
a reasonable alternative approach based on the concept of relative ageing to handle the crossing hazard rates problem. They have defined some stochastic orderings (some of which are originally defined by Kalashnikov and Rachev~\cite{kr6}) based on the concept of relative ageing.
\\\hspace*{0.2 in} Stochastic orderings and different ageing classes are extensively studied in reliability theory. Stochastic orderings are used to compare two life distributions from different aspects.   
Many different types of stochastic orderings have been developed (see Shaked and Shanthikumar~\cite{shak1}). On the other hand, positive ageing means an older system has shorter remaining
lifetime than a younger one in some stochastic sense. Many different types of life distributions are characterized by their ageing properties. It has been observed that
some stochastic orderings which compare probability
distributions with the exponential distribution are found to be very useful to understand the phenomenon
of ageing (see Barlow and Proschan~\cite{bp1}, Deshpande \emph{et al.}~(\cite{dks1}, \cite{dsbj1}), Kochar and Wiens~\cite{kw6}, 
Lai and Xie~\cite{lx6}, and the references therein). 
We introduce some new generalized stochastic orderings (in the spirit of relative ageing) which 
compare probability distributions with the exponential distribution.
These orderings may be useful to realize the phenomenon of positive ageing classes from different angles and also may be helpful to get guidance for the crossing hazard rates and/or
crossing mean residual lives problem.   
\\\hspace*{0.2 in}For any absolutely continuous nonnegative random variable $X$, let the probability density function be denoted by 
$f_X(\cdot)$, the cumulative distribution function by $F_X(\cdot)$ and the survival function by $\overline{F}_X(\cdot)=1-F_X(\cdot)$.
 Let us write
$$\overline{T}_{X,0}(x)=f_X(x),$$
and
\begin{equation}
\overline{T}_{X,s}(x)=\frac{\int_x^\infty \overline{T}_{X,s-1}(t)dt}{\tilde\mu_{X,s-1}},\label{eqn1-1}
\end{equation}
for $s=1,2,\ldots$, where
$$\tilde\mu_{X,s}=\int_0^\infty \overline{T}_{X,s}(t)dt,$$
$s=0,1,2,\ldots.$ We assume $\tilde \mu_{X,s}$ to be finite for all $s=0,1,2,\ldots.$ 
We denote the random variable corresponding to the survival function $\overline{T}_{X,s}(\cdot)$ by $X_s$. Clearly, $X_1\equiv X$. We further define, for $s=1,2,\dots$,
\begin{eqnarray}
\La_{X,s}(\cdot)=-\log \ov T_{X,s}(\cdot),
\end{eqnarray}
\begin{eqnarray*}
r_{X,s}(x)&=&\frac{\overline{T}_{X,s-1}(x)}{\int_x^\infty \overline{T}_{X,s-1}(t)dt}\\
&=&\frac{\overline{T}_{X,s-1}(x)}{\tilde \mu_{X,s-1}\overline{T}_{X,s}(x)},
\end{eqnarray*}
and 
\begin{eqnarray*}
 \mu_{X,s}(x)&=&\frac{\int_x^\infty \overline{T}_{X,s}(t)dt}{\overline{T}_{X,s}(x)},
\end{eqnarray*}

where $\La_{X,s}(\cdot)$, $r_{X_s}(\cdot)$ and $\mu_{X,s}(\cdot)$, respectively, represent the cumulative hazard function, the failure rate function and the mean residual life function corresponding to $X_s$.
%For $s=1$, $r_{X,1}(x)$ is the failure rate of $X$, defined as the ratio of
%the probability density to the survival, whereas, for $s=2$, $r_{X,2}(t)$ is the reciprocal
%of the mean residual life $E(X_t)=E\left(X-t|X\ge t\right)$. 
Note that, for $s=1,2,\dots$,
$$\mu_{X,s}(0)=\tilde \mu_{X,s},$$  and, for $s=2,3,\dots$,
\begin{eqnarray}\label{mre5}
r_{X,s}(x)=\frac{1}{\mu_{X,s-1}(x)}.
\end{eqnarray}
%It is worth to mention here that $$\ov T_{X,0}(0)=f_X(0),\;\ov T_{Y,0}(0)=f_Y(0),\;\tilde \mu_{X,0}=\tilde \mu_{Y,0}=1,$$ and, for $s=2,3,\dots$, $$\ov T_{X,s-1}(0)=\ov T_{Y,s-1}(0)=1.$$
%%%%%%%%%%%%%
\\\hspace*{0.2 in}Throughout the paper, increasing and decreasing properties are not used in strict sense. For any differentiable function $k(\cdot)$, we write $k'(t)$ to denote the first 
derivative of $k(t)$ with respect to $t$.
\\\hspace*{0.2 in}The following well known definitions may be obtained in Fagiuoli and Pellerey~\cite{fp6}.
\begin{definition} \label{def1-1} For $s=1,2,\ldots$, $X$ is said to be 
\begin{enumerate}
\item[($i$)] s-IFR if $r_{X,s}(x)$ is increasing in $x\ge 0$;
\item[($ii$)] s-IFRA if $\frac{1}{x}\int_0^x r_{X,s}(t)dt$ is increasing in $x > 0$;
\item[($iii$)] s-NBU if $\overline{T}_{X,s}(x+t)\le \overline{T}_{X,s}(x).\overline{T}_{X,s}(t)$ 
for all $x,t\ge 0$;
\item[($iv$)] s-NBUFR if $r_{X,s}(0)\le r_{X,s}(x)$ for all $x\ge 0;$
\item[($v$)] s-NBAFR if $r_{X,s}(0)\le \frac{1}{x}\int_0^x r_{X,s}(x)$ for all $x> 0.$
\end{enumerate}
\end{definition}
It is easy to verify that each of the following equivalence relations holds:\\\\
\begin{tabular}{lll}
 1-IFR $\Leftrightarrow$ IFR,& 2-IFR $\Leftrightarrow$ DMRL,&
3-IFR $\Leftrightarrow$ DVRL,\\
1-IFRA $\Leftrightarrow$ IFRA,& 2-IFRA $\Leftrightarrow$ DMRLHA,&
1-NBU $\Leftrightarrow$ NBU,\\
 1-NBUFR $\Leftrightarrow$ NBUFR,&
2-NBUFR $\Leftrightarrow$ NBUE, &3-NBUFR $\Leftrightarrow$ NDVRL,\\
%1-NBUCX$\Leftrightarrow$NBUC;&1-NBUCV$\Leftrightarrow$NBU(2);&
1-NBAFR$\Leftrightarrow$NBAFR,&2-NBAFR$\Leftrightarrow$HNBUE.&
\end{tabular}
\\\hspace*{0.2 in}For the definitions of 
 IFR (Increasing
in Failure Rate), IFRA (Increasing in Failure Rate Average), NBU (New Better than Used),
DMRL (Decreasing in Mean Residual Life) and NBUE (New Better than Used in Expectation) classes
one may refer to Bryson and Siddiqui~\cite{bs6}, and Barlow and Proschan~\cite{bp1};
DVRL (Decreasing in Variance Residual Life) and NDVRL (Net DVRL)
classes are discussed in Launer~\cite{l6};
DMRLHA (Decreasing Mean Residual Life in Harmonic Average)
and NBUFR (New Better than Used in Failure Rate) classes are
studied by Deshpande et al.~\cite{dks1}; NBAFR (New Better Than Used in Failure Rate Average) is due to Loh~\cite{lw1}, whereas
HNBUE (Harmonically New Better than Used in Expectation) is discussed in Klefsj$\ddot{\rm{o}}$~\cite{kle1}. Similarly, the negative ageing notions, namely, DFR, DFRA, NWU, NWUFR,
NWAFR, etc. are also found in the literature. 
\\\hspace*{0.2 in} Let ${\cal F}$ be the class of distribution functions $F:[0,\infty)\longrightarrow [0,1]$
with $F(0)=0$. We assume that all $F(\in {\cal F})$ have their finite generalized means $\wt \mu_{X,s}$, and are strictly
increasing on their support. If $F$ is not strictly increasing, we take
the inverse as 
$$F^{-1}(y)=\inf\{x:F(x)\ge y\}.$$
%%%%%%%%%%%%%%%%%%%%%%%%%%%%%%%%%%%%%%%%%%%%%%%%%%%%%%%%%%%%%%%%%%%%%%%%%%%%%%%%%%%%%%%%%%%%%%%% 
\hspace*{.2in} A function $f(\cdot)$ is called star-shaped (resp. antistar-shaped) if $f(x)/x$ is increasing (resp. decreasing) in $x>0$.
On the other hand, it is called super-additive (resp. sub-additive) if, for all $x,y$, $f(x+y)\geq(resp. \leq~)f(x)+f(y)$.
\\\hspace*{.2in} Let $Y$ be another absolutely continuous nonnegative random variable with respective generalized functions (analogous to the one
defined above for $X$) $\overline{T}_{Y,s}(\cdot),\;\wt \mu_{Y,s}$, $\La_{Y,s}(\cdot)$, $r_{Y,s}(\cdot)$ and $\mu_{Y,s}(\cdot)$. The random variable corresponding to the survival function 
$\ov T_{Y,s}(\cdot)$ is denoted by $Y_s$. For the sake of simplicity we write, for $s=1,2,\dots,$
$$\Phi^s_{X,Y}\equiv \Lambda_{Y,s}(X_s).$$
\hspace*{0.2 in}In this note, we define some new generalized stochastic orderings, and some of their properties are also studied. In Sections $2$, $3$, $4$, $5$ and $6$, we discuss s-IFR(R), s-IFRA(R),
s-NBU(R), s-NBUFR(R) and s-NBAFR(R) orderings, respectively.
We write `R' within parenthesis to mean that this ordering has been generated based on the concept of {\bf R}elative ageing.
Some characterizations of these orderings are discussed. We show that these orderings are scale and base invariant.
Inter-relations among these orderings have also been discussed. We build a bridge by which 
these orderings could connect to the generalized ageing classes, and vice versa.
\section{s-IFR(R) Ordering}\label{sec2}
%%%%%%%%%%%%%%%%%%%%%%%%%%%%%%%%%%%%%%%%%%%%%%%%%%%%  Definition   %%%%%%%%%%%%%%%%%%%%%%%%%%%%%%%%%%%%%%%
\hspace*{0.2 in}In this section we define and study s-IFR(R) ordering. 
 This ordering interprets that ratio of the hazard rates
 of $X_s$ and $Y_s$ is increasing. This means that $X_s$ ages faster than $Y_s$. We begin with the following definition.
\begin{definition} \label{def2*1} For any positive integer $s$, $X$ (or its distribution $F_X$) is said to be more $s$-IFR(R) than  
$Y$ (or its distribution $F_Y$) (written as $X\leq_{s-IFR(R)}Y$) if the random variable 
$\Phi^s_{X,Y}\;\text{has an IFR distribution}$.\hfill $\Box$
\end{definition}
%%%%%%%%%%%%%%%%%%%%%%%%%%%%%%%%%%%%%%%%%%%%%%%%%%%%%%%%%%%% Remark %%%%%%%%%%%%%%%%%%%%%%%%%%%%%%%%%%%%%%%%%%%%%%%%%
\begin{remark}
For $s=1$, Definition \ref{def2*1} gives $X\leq_{c}Y$, as discussed in Sengupta and Deshpande~\cite{sd2}.\hfill $\Box$
\end{remark}
\hspace*{0.2 in}The following lemma may be obtained in Marshall and Olkin~(\cite{mo1}, Section $21(f)$, pp. 699-700). 
\begin{lem}\label{l2-1}
 Let $f(\cdot)$ and $g(\cdot)$ be two real-valued continuous functions, and $\zeta(\cdot)$ be a strictly increasing (resp. decreasing) and continuous function defined on the range of $f$ and $g$.
 Then, for any real number $c>0$, $f(x)-cg(x)$ and $\zeta (f(x))-\zeta (cg(x))$ have sign change property in the same (resp. reverse) order, as $x$ traverses from left to right.\hfill $\Box$
\end{lem}
%%%%%%%%%%%%%%%%%%%%%%%%%%%%%%%%%%%%%%%%%%%%%%  Proposition %%%%%%%%%%%%%%%%%%%%%%%%%%%%%%%%%%%%%%%%%%%%%%%%%
\hspace*{0.2 in}In the following two propositions, we give some equivalent representations of the s-IFR(R) ordering. The second proposition 
can easily be verified by using Lemma \ref{l2-1} or Proposition~2.C.8 of Marshall and Olkin~\cite{mo1}, and Proposition~\ref{pp61}(i).
\begin{pro}\label{pp61}
 For $s=2,3,\dots$, Definition \ref{def2*1} can equivalently be written in one of the following forms: 
 \begin{enumerate}
 \item [$(i)$] $\La_{X,s}\left(\La^{-1}_{Y,s}(x)\right)$ is convex in $x\geq 0.$
 \item [$(ii)$] $\frac{r_{X,s}(x)}{r_{Y,s}(x)}\;\text{is increasing in}\;x\geq0.$
 \item [$(iii)$] $\frac{\mu_{X,s-1}(x)}{\mu_{Y,s-1}(x)}\;\text{is decreasing in}\;x\geq0.$
 \item [$(iv)$] $\Lambda_{X,s}( Y_s)\;\text{has a DFR distribution.}$
  \end{enumerate}
\end{pro}
{\bf Proof:} We have 
\begin{eqnarray}
\La_{X,s}\left(\La^{-1}_{Y,s}(x)\right)&=&-\log\left(\ov T_{X,s}\ov T^{-1}_{Y,s}\left(e^{-x}\right)\right)\nonumber
\\&=&-\log \ov F_{\Phi^s_{X,Y}}(x).\label{rre14}
\end{eqnarray}
Thus, $(i)$ follows from Definition \ref{def2*1}, and conversely.
Again, Definition \ref{def2*1} can equivalently be written as
\begin{eqnarray}\label{rre15}
r_{\Phi^s_{X,Y}}(x)=\left(\frac{\tilde \mu_{Y,s-1}}{\tilde \mu_{X,s-1}}\right)\left(\frac{\ov T_{X,s-1}\ov T^{-1}_{Y,s}\left(e^{-x}\right)}{\ov T_{X,s}\ov T^{-1}_{Y,s}\left(e^{-x}\right)}\right)\left(\frac{e^{-x}}{\ov T_{Y,s-1}\ov T^{-1}_{Y,s}\left(e^{-x}\right)}\right)
\end{eqnarray}
is increasing in $x\geq0$, which holds if, and only if,
$$\left(\frac{\tilde \mu_{Y,s-1}}{\tilde \mu_{X,s-1}}\right)\left(\frac{\ov T_{X,s-1}\ov T^{-1}_{Y,s}\left(u\right)}{\ov T_{X,s}\ov T^{-1}_{Y,s}\left(u\right)}\right)\left(\frac{u}{\ov T_{Y,s-1}\ov T^{-1}_{Y,s}\left({u}\right)}\right)\;\text{is decreasing in}\;u\in(0,1],$$
%or equivalently,
%$$\left(\frac{\tilde \mu_{Y,s-1}}{\tilde \mu_{X,s-1}}\right)\left(\frac{\ov T_{X,s-1}\left(x\right)}{\ov T_{X,s}\left(x\right)}\right)\left(\frac{\ov T_{Y,s}(x)}{\ov T_{Y,s-1}\left({x}\right)}\right)\;\text{is increasing in}\;x\geq 0,$$
or equivalently,
$$ \frac{r_{X,s}(x)}{r_{Y,s}(x)}\;\text{is increasing in}\;x\geq0.$$ This gives the equivalence of Definition \ref{def2*1} and $(ii)$. The one-to-one connection between $(ii)$ and $(iii)$
follows from (\ref{mre5}). 
 Note that $(iv)$ holds if, and only if,
$$\frac{r_{Y,s}(x)}{r_{X,s}(x)}\;\text{is decreasing in}\;x\geq0,$$ which is equivalent to $(ii)$.\hfill $\Box$
\begin{remark}
 The equivalences of $(i)$, $(ii)$ and $(iv)$ given in Proposition~\ref{pp61} are also true for $s=1$.\hfill $\Box$
\end{remark}
%%%%%%%%%%%%%%%%%%%%%%%%%%%%%%%%%%%%%%%%%%%%%%%%%%%%%%%%%%%%%%%%%%%%%%%%%%%%%%%%%%%%%%%%%%%%%%%%%%%%%%%%%%%%%%%%%%%%%%%%%%%%%
\begin{pro}
 Definition \ref{def2*1} can equivalently be written in one of the following forms:
 \begin{enumerate}
 %\item [$(i)$] $\ov T^{-1}_{Y,s}\ov T_{X,s}(x)$ is convex in
  \item [$(i)$] For any real numbers $a$ and $b$, $\La_{X,s}\La^{-1}_{Y,s}(x)-(ax+b)$ changes sign at most twice, and if the change of signs occurs twice, they are in the order $+,-,+$, as $x$ traverses from $0$~to~$\infty$.
   \item [$(ii)$] For any real numbers $a$ and $b$, $\La^{-1}_{Y,s}(x)-\La^{-1}_{X,s}(ax+b)$ changes sign at most twice, and if the change of signs occurs twice, they are in the order $+,-,+$,  as $x$ traverses from $0$~to~$\infty$.
    \item [$(iii)$] For any real numbers $a$ and $b$, $\La^{-1}_{Y,s}(ax+b)-\La^{-1}_{X,s}(x)$ changes sign at most twice, and if the change of signs occurs twice, they are in the order $+,-,+$,  as $x$ traverses from $0$~to~$\infty$.
     \item [$(iv)$] For any real numbers $a$ and $b$, $\La^{-1}_{X,s}(x)-\La^{-1}_{Y,s}(ax+b)$ changes sign at most twice, and if the change of signs occurs twice, they are in the order $-,+,-$,  as $x$ traverses from $0$~to~$\infty$.
      \item [$(v)$] For any real numbers $a$ and $b$, $\La_{Y,s}\La^{-1}_{X,s}(x)-(ax+b)$ changes sign at most twice, and if the change of signs occurs twice, they are in the order $-,+,-$,  as $x$ traverses from $0$~to~$\infty$.
      \item [$(vi)$] $\La_{Y,s}\La^{-1}_{X,s}(x)$ is concave in $x>0$.\hfill $\Box$
 \end{enumerate}
\end{pro}
%%%%%%%%%%%%%%%%%%%%%%%%%%%%%%%%%%%%%%%%%%%%%%%%%%%%%%%%%%%%%%%%%%%%%%   lemma %%%%%%%%%%%%%%%%%%%%%%%%%%%%%%%%%%%%%%%%%%%%%%%%%%%%%%%%%%%%%%%%%%%%%%%%%%%%%%%%%%%%%%%%%%%%%%%%%%%%%%%%%%
\hspace*{0.2 in}Below we state two lemmas which will be used in proving the upcoming theorem. The proofs are omitted.
\begin{lem}\label{lem2-6}
 Let $f(\cdot)$ be a nonnegative, increasing, and convex function. Then $f^{-1}(\cdot)$ is concave.\hfill $\Box$
\end{lem}
%%%%%%%%%%%%%%%%%%%%%%%%%%%%%%%%%%%%%%%%%%%%%%%%%%%%%%%%%%%%%%%%%%%%%%%%%%%%%%%%%%%%%%%%%%%%%%%%%%%%%%%%%%%%%%%%%%%%%%%%%%%%%%%%%%%%%%%%%%%%%%%%%%%%%%%%%%%%%%%%%%%%%%%%
\begin{lem}\label{lem2-5}
 Let $f(\cdot)$ and $g(\cdot)$ be two nonnegative, increasing, and convex functions. Then $f(g(\cdot))$ is convex.\hfill $\Box$
\end{lem}
%%%%%%%%%%%%%%%%%%%%%%%%%%%%%%%%%%%%%%%%%%%%%%%%%%%%%%%%%%%%%%%%%%%%%%%%%%%%%%%%%%   Theorem  %%%%%%%%%%%%%%%%%%%%%%%%%%%%%%%%%%%%%%%%%%%%%%%%%%%%%%%%%%%%%%%%%%%%%%%%%%%%%%%%%%%%%%%%%%%%%%%
\hspace*{0.2 in}The following theorem shows some properties of the s-IFR(R) ordering.
\begin{thm}\label{th2*1}
For any positive integer $s$,
\begin{enumerate}
%\vspace*{2 in}
\item [$(i)$] $X\leq_{s-IFR(R)}X$.
 \item [$(ii)$] $X\leq_{s-IFR(R)}Y$ and $Y\leq_{s-IFR(R)}X$ hold simultaneously if, and only if, $\La_{X,s}(x)=\theta\La_{Y,s}(x)$, for some $\theta>0$ and for all $x\geq 0.$
 \item [$(iii)$] If $X\leq_{s-IFR(R)}Y$ and $Y\leq_{s-IFR(R)}Z$ then $X\leq_{s-IFR(R)}Z$.
\end{enumerate}
\end{thm}
{\bf Proof:} The proof of $(i)$ is trivial. Now,
$X\leq_{s-IFR(R)}Y$ gives that $$\La_{X,s}\left(\La^{-1}_{Y,s}(x)\right)\;\text{is convex in}\;x\geq0,$$ which, by Lemma \ref{lem2-6}, 
reduces to the fact that 
\begin{equation}\label{aa}
 \La_{Y,s}\left(\La^{-1}_{X,s}(x)\right)\;{\rm is\; concave}.
\end{equation}
 Further, $Y\leq_{s-IFR(R)}X$ gives that
\begin{equation}\label{ab}
 \La_{Y,s}\left(\La^{-1}_{X,s}(x)\right)\;{\rm is\; convex}.
\end{equation}
Combining (\ref{aa}) and (\ref{ab}), we get
$$\La_{Y,s}\left(\La^{-1}_{X,s}(x)\right)=\frac{x}{\theta} ,$$
for some constant $\theta\;(>0)$. 
%Now, by evaluating the above expression at $x=0$, we get $\alpha=0$. Hence, we have 
%$$\La_{Y,s}\left(\La^{-1}_{X,s}(x)\right)=\frac{x}{\theta },$$
Thus, $\La_{X,s}(x)=\theta\La_{Y,s}(x)$, and hence $(ii)$ is proved.
On using Lemma \ref{lem2-5}, one can easily check that $(iii)$ holds.\hfill $\Box$
%%%%%%%%%%%%%%%%%%%%%%%%%%%%%%%%%%%%%%%%%%%%%%%%%%%%%%%%%%%%%%%%%%%%%%%%%%%%%%%%%%%%%%%%%%%%%%%  Theorem %%%%%%%%%%%%%%%%%%%%%%%%%%%%%%%%%%%%%%%%%%%%%%%%%%%%%%%%%%%%%%%%%%%%%%%%%%%%%%
\\\hspace*{.2in} The following lemma can be easily verified.
\begin{lem}\label{lem2-3} Let $X\sim \ov F_{X}(x)=e^{-\lambda x}$. Then, for $s=1,2,\ldots$,
\begin{enumerate}
\item[($i$)] $r_{X,s}(x)=\lambda $;
\item[($ii$)] $\overline{T}_{X,s}(x)=e^{-\lambda x}.$\hfill $\Box$
\end{enumerate}
\end{lem}
%%%%%%%%%%%%%%%%%%%%%%%%%%%%%%%%%%%%%%%%%%%%%%%%%%%%%%%%%%%%%%%%%%%%   Theorem   %%%%%%%%%%%%%%%%%%%%%%%%%%%%%%%%%%%%%%%%%%%%%%%%%%%%%%%%%%%%%%%%%%%%%%%%%%%%%%%%%%%%%%%%%%%%%%%%%%%%%%%%55555
%\begin{thm}\label{th2*1} The relationship $X\leq_{s-IFR(R)}Y$ is a partial
%ordering of the equivalence classes of $\cal{F}$.
%\end{thm}
%%%%%%%%%%%%%%%%%%%%%%%%%%%%%%%%%%%%%%%%%%%%%%%%%%%%%%%%%%%%    Theorem  %%%%%%%%%%%%%%%%%%%%%%%%%%%%%%%%%%%%%%%%%%%%%%%%%%%%%%%%%%%%%%%%%%
\hspace*{0.2 in}The following theorem shows that a random variable $X$ has an s-IFR distribution if, and only if, $X$ is smaller than exponential distribution in s-IFR(R) ordering. The proof follows from 
Lemma~\ref{lem2-3}.
\begin{thm}
 If $\ov F_Y(x)=e^{-\la x}$, then $X\leq_{s-IFR(R)}Y$ if, and only if, $X$ is s-IFR.\hfill $\Box$
\end{thm}
%%%%%%%%%%%%%%%%%%%%%%%%%%%%%%  Characterization theorem  %%%%%%%%%%%%%%%%%%%%%%%%%%%%%%%%%%%%%%%%%%%%%%%%%%%%%%%%%%%%%%%%%%%%%%%%%%%%%%%%%%%%%%%%%%%%%%%%%%%%%%%%%%%%%%%%%%%%%%%%%%%%%%%%%
\hspace*{0.2 in} Below we give a lemma which will be used in proving the upcoming theorem, and can be proved using Principle of Mathematical Induction.
\begin{lem}\label{lehk}
  For any real numbers $a\;(>0)$ and $b$, and for $s=1,2,\dots$, 
 \begin{enumerate}
 \item [$(i)$] $\overline{T}_{aX+b,s}(x)=\overline{T}_{X,s}\left(\frac{x-b}{a}\right)$.
  \item [$(ii)$] $\tilde \mu_{aX+b,s}=a\tilde \mu_{X,s}$.
  \item [$(iii)$] $r_{aX+b,s}(x)=\frac{1}{a}r_{X,s}\left(\frac{x-b}{a}\right)$.\hfill $\Box$ 
 \end{enumerate}
\end{lem}
\hspace*{0.2 in}The following theorem shows that s-IFR(R) ordering is location and scale invariant. 
\begin{thm}
 $X\leq_{s-IFR(R)}Y$ if, and only if, $\left(aX+b\right)\leq_{s-IFR(R)}\left(aY+b\right)$, for any real numbers $a\;(>0)$ and $b$.\hfill $\Box$
\end{thm}
{\bf Proof:} Let $\La_{aX+b,s}(\cdot)$ and $\La_{aY+b,s}(\cdot)$ be the cumulative hazard rate functions of $aX+b$ and $aY+b$, respectively. 
Then, on using Lemma~\ref{lehk}, we have, for all $x\geq0$,
%Then, we have, for all $x\geq0$, 
 \begin{eqnarray}
 \La_{aX+b,s}\left(\La_{aY+b,s}\right)^{-1}(x)&=&-\log\left(\ov T_{aX+b,s}\ov T^{-1}_{aY+b,s}\left(e^{-x}\right)\right)\nonumber
 \\&=&-\log\left[\ov T_{X,s}\left(\frac{\ov T^{-1}_{aY+b,s}\left(e^{-x}\right)-b}{a}\right)\right]\nonumber
 \\&=&-\log\left[\ov T_{X,s}\left(\frac{b+a\ov T^{-1}_{Y,s}\left(e^{-x}\right)-b}{a}\right)\right]\nonumber
 %\\&=&-\log\left(\ov T_{X,s}\ov T^{-1}_{Y,s}\left(e^{-x}\right)\right)\nonumber
\\ &=&\La_{X,s}\La^{-1}_{Y,s}(x).\label{rre19}
 \end{eqnarray}
 Thus, the result follows from Proposition \ref{pp61}(i).\hfill $\Box$
% \hspace*{0.2 in}In the following theorem, we prove that s-IFR(R) ordering is preserved under strictly increasing positive transformation. 
% \begin{thm}\label{thm2*2}
%  $X\leq_{s-IFR(R)}Y$ if, and only if, $g(X_s)\leq_{1-IFR(R)}g(Y_s)$ for every strictly increasing positive function $g(\cdot)$.
% \end{thm}
% {\bf Proof:} Let $\La^g_{X,s}(\cdot)$ be the cumulative hazard function of $g(X_s)$. Note that, for all $x\geq0$,
% \begin{eqnarray}
% \La^g_{X,s}(\La^g_{Y,s})^{-1}(x)=\La_{X,s}\La^{-1}_{Y,s}(x).\label{rre19}
% \end{eqnarray}
% Thus, the result follows from Proposition \ref{pp61}(a).\hfill $\Box$
% \\\hspace*{0.2 in}The following corollary is an immediate consequence of the above theorem.
% \begin{coro}
% \begin{enumerate}
%  \item [$(a)$] $X\leq_{s-IFR(R)}Y$ if, and only if, $pX\leq_{s-IFR(R)}pY$ for all $p>0$.
%  \item [$(b)$] $X\leq_{s-IFR(R)}Y$ if, and only if, $X^r\leq_{s-IFR(R)}Y^r$ for all $r>0$.\hfill $\Box$
%  \end{enumerate}
% \end{coro}
%%%%%%%%%%%%%%%%%%%%%%%%%%%%%%%%%%%%%%%%%%%%%%%%%%%%%%%%%%%%%%%%%%%%%%%%%%%%%%%%%%%%%%%%%%%%%%  s-AFA  %%%%%%%%%%%%%%%%%%%%%%%%%%%%%%%%%%%%%%%%%%%%%%%%%%%
\section{s-IFRA(R) Ordering}\label{sec3}
\hspace*{0.2 in}In this section we discuss s-IFRA(R) ordering. The interior scenario of s-IFRA(R) ordering is that ratio of the cumulative hazard rates of $X_s$ and $Y_s$ is increasing.
\begin{definition} \label{def3*1} For any positive integer $s$, $X$ (or its distribution $F_X$) is said to be more $s$-IFRA(R) than  
$Y$ (or its distribution $F_Y$) (written as $X\leq_{s-IFRA(R)}Y$) if the random variable 
$\Phi^s_{X,Y}\;\text{has an IFRA distribution}$.\hfill $\Box$
\end{definition}
\begin{remark}
For $s=1$, Definition \ref{def3*1} gives $X\leq_{*}Y$, as discussed in Sengupta and Deshpande~\cite{sd2}.\hfill $\Box$
\end{remark}
%%%%%%%%%%%%%%%%%%%%%%%%%%%%%%%%%%%%%%%%%%%%%%%%%%%%%%%%%%%%%%%%%%%%%%%%%%%%%%%%%%%%%%%%%%%%%%%  Proposition   %%%%%%%%%%%%%%%%%%%%%%%%%%%%%%%%%%%%%%%%%%%%%%%%%%%%%%%%%%%%%%%%%%%%%
\hspace*{0.2 in}Some equivalent representations of s-IFRA(R) ordering are given in the following two propositions. The second proposition can easily be proved by Lemma~\ref{l2-1} and
Proposition~\ref{ar6}(i).
\begin{pro}\label{ar6}
 Definition \ref{def3*1} can equivalently be written in one of the following forms:
 \begin{enumerate}
 \item [$(i)$] $\La_{X,s}\left(\La^{-1}_{Y,s}(x)\right)$ is star-shaped in $x> 0.$
 \item [$(ii)$] $\frac{\La_{X,s}(x)}{\La_{Y,s}(x)}\;\text{is increasing in}\;x>0.$
% \item [$(c)$] $\frac{\mu_{X,s-1}(x)}{\mu_{Y,s-1}(x)}\;\text{is decreasing in}\;x\geq0.$
 \item [$(iii)$] $\Lambda_{X,s}(Y_s)\;\text{has a DFRA distribution.}$
  \end{enumerate}
\end{pro}
{\bf Proof:} On using (\ref{rre14}), the equivalence of Definition~\ref{def3*1} and $(i)$ follows. Note that $(i)$ can equivalently be written as

$$\frac{\La_{X,s}\left(\La^{-1}_{Y,s}(x)\right)}{x}\;\text{is increasing in}\;x>0,$$
or equivalently,
$$\frac{\La_{X,s}(x)}{\La_{Y,s}(x)}\;\text{is increasing in}\;x>0.$$ Thus, the equivalence of $(i)$ and $(ii)$ is proved. The one-to-one connection between $(ii)$ and $(iii)$
can be proved in the same line as is done in $(ii)$ above.\hfill $\Box$
%$$\frac{\tilde \mu_{Y,s-1}}{\tilde \mu_{X,s-1}}\left(\frac{\ov T_{X,s-1}\ov T^{-1}_{Y,s}\left(e^{-x}\right)}{\ov T_{X,s}\ov T^{-1}_{Y,s}\left(e^{-x}\right)}\right)\left(\frac{e^{-x}}{\ov T_{Y,s-1}\ov T^{-1}_{Y,s}\left(e^{-x}\right)}\right)\geq \frac{\La_{X,s}\left(\La^{-1}_{Y,s}(x)\right)}{x},$$
\begin{pro}\label{pn3-1}
 Definition \ref{def3*1} can equivalently be written in one of the following forms:
 \begin{enumerate}
  \item [$(i)$] For any real number $a$, $\La_{X,s}\La^{-1}_{Y,s}(x)-ax$ changes sign at most once, and if the change of sign does occur, it is in the order $-,+$, as $x$ traverses from $0$ to $\infty$.
   \item [$(ii)$] For any real number $a$, $\La^{-1}_{Y,s}(x)-\La^{-1}_{X,s}(ax)$ changes sign at most once, and if the change of sign does occur, it is in the order $-,+$,  as $x$ traverses from $0$ to $\infty$.
    \item [$(iii)$] For any real number $a$, $\La^{-1}_{Y,s}(ax)-\La^{-1}_{X,s}(x)$ changes sign at most once, and if the change of sign does occur, it is in the order $-,+$,  as $x$ traverses from $0$ to $\infty$.
     \item [$(iv)$] For any real number $a$, $\La^{-1}_{X,s}(x)-\La^{-1}_{Y,s}(ax)$ changes sign at most once, and if the change of sign does occur, it is in the order $+,-$,  as $x$ traverses from $0$ to $\infty$.
      \item [$(v)$] For any real number $a$, $\La_{Y,s}\La^{-1}_{X,s}(x)-ax$ changes sign at most once, and if the change of sign does occur, it is in the order $+,-$,  as $x$ traverses from $0$ to $\infty$.
      \item [$(vi)$] $\La_{Y,s}\La^{-1}_{X,s}(x)$ is antistar-shaped in $x>0$.\hfill $\Box$
 \end{enumerate}
\end{pro}
%%%%%%%%%%%%%%%%%%%%%%%%%%%%%%%%%%%%%%%%%%%%%%%%%%%%%%%%%%%%%%%%%%%%   Lemma  %%%%%%%%%%%%%%%%%%%%%%%%%%%%%%%%%%%%%%%%%%%%%%%%%%%%%%%%%%%%%%%%%%%%%%%%%%%%%%%%%%%%%%%%%%%%%%%%%%%%%%%%55555
\hspace*{0.2 in} Before going to the next theorem we give two lemmas without proof.
\begin{lem}\label{lem3-2}
 Let $f(\cdot)$ be a nonnegative, increasing, and star-shaped function. Then $f^{-1}(\cdot)$ is antistar-shaped.\hfill$\Box$
\end{lem}
% {\bf Proof:} Given that $f(x)/x$ is increasing. This means
% \begin{equation}\label{ac}
%  xf'(x)\geq f(x).
% \end{equation}
% Now 
% \begin{eqnarray*}
%  \frac{d}{dx}\left(\frac{f^{-1}(x)}{x}\right)&=&\frac{x}{f'(f^{-1}(x))}-f^{-1}(x)\\
%  &=&\frac{f(f^{-1}(x))}{f'(f^{-1}(x))}-f^{-1}(x)\\
%  &\leq &f^{-1}(x)-f^{-1}(x)\\
%  &=&0.
% \end{eqnarray*}
% The above inequality follows from (\ref{ac}).\hfill$\Box$
%%%%%%%%%%%%%%%%%%%%%%%%%%%%%%%%%%%%%%%%%%%%%%%%%%%%%%%%%%%%%%%%  Lemma %%%%%%%%%%%%%%%%%%%%%%%%%%%%%%%%%%%%%%%%%%%%%%%%%%%%%%%%%%%%%%%%%%%%%%%%%%%%%%%%%%%%%%%%%%%%%%%%%%%%%%%%%%%%%%%
\begin{lem}\label{lem3-1}
 Let $f(\cdot)$ and $g(\cdot)$ be two nonnegative, increasing, and star-shaped functions. Then $f(g(\cdot))$ is star-shaped.\hfill$\Box$
\end{lem}
%%%%%%%%%%%%%%%%%%%%%%%%%%%%%%%%%%%%%%%%%%%%%%%%%%%%%%%%%%%%%%%%%%%%%%%%%%%%%%%%%%%%%%%%%%%%%  Theorem  %%%%%%%%%%%%%%%%%%%%%%%%%%%%%%%%%%%%%%%%%%%%%%%%%%%%%%%%%%%%%%%%%%%%%%%%%%%
\hspace*{0.2 in}Some properties of the s-IFRA(R) ordering are discussed in the following theorem.
\begin{thm}\label{th3*1}
For any positive integer $s$,
\begin{enumerate}
%\vspace*{2 in}
\item [$(i)$] $X\leq_{s-IFRA(R)}X$.
 \item [$(ii)$] $X\leq_{s-IFRA(R)}Y$ and $Y\leq_{s-IFRA(R)}X$ hold simultaneously if, and only if, $\La_{X,s}(x)=\theta\La_{Y,s}(x)$, for some $\theta>0$ and for all $x\geq 0.$
 \item [$(iii)$] If $X\leq_{s-IFRA(R)}Y$ and $Y\leq_{s-IFRA(R)}Z$ then $X\leq_{s-IFRA(R)}Z$.
\end{enumerate}
\end{thm}
{\bf Proof:} The proof of $(i)$ is trivial. To prove $(ii)$ we proceed as follows.
\\ $X\leq_{s-IFRA(R)}Y$ gives that $$\La_{X,s}(\La^{-1}_{Y,s}(\cdot))\;\text{is star-shaped},$$ which, by Lemma~\ref{lem3-2},
reduces to the fact that  $$\La_{Y,s}(\La^{-1}_{X,s}(\cdot))\;\text{is antistar-shaped}.$$ Further, 
$Y\leq_{s-IFRA(R)}X$ gives that $$\La_{Y,s}(\La^{-1}_{X,s}(\cdot))\;\text{is star-shaped}.$$ Combining the two, we have
$$\La_{Y,s}(\La^{-1}_{X,s}(x))=\frac{x}{\theta},$$
for some constant $\theta\;(>0)$. Thus, we have $\La_{X,s}(x)=\theta\La_{Y,s}(x)$, and hence $(ii)$ is proved.
Again, by Lemma \ref{lem3-1}, $(iii)$ holds.\hfill$\Box$
%%%%%%%%%%%%%%%%%%%%%%%%%%%%%%%%%%%%%%%%%%%%%%%%%%%%%%%%%%%%    Theorem  %%%%%%%%%%%%%%%%%%%%%%%%%%%%%%%%%%%%%%%%%%%%%%%%%%%%%%%%%%%%%%%%%%
\\\hspace*{0.2 in}The following theorem is a bridge between s-IFRA(R) ordering and s-IFRA ageing class. The proof follows from Lemma~\ref{lem2-3}.
\begin{thm}
 If $\ov F_Y(x)=e^{-\la x}$, then $X\leq_{s-IFRA(R)}Y$ if, and only if, $X$ is s-IFRA.\hfill$\Box$
\end{thm}
\hspace*{0.2 in}Since, every IFR distribution is an IFRA distribution, we have the following theorem.
\begin{thm}
 If $X\leq_{s-IFR(R)}Y$ then $X\leq_{s-IFRA(R)}Y.$\hfill$\Box$
\end{thm}
% %%%%%%%%%%%%%%%%%%%%%%%%%%%%%%%%%%%%%%%%%%%%%%%%%%%%%%%%%%%%%%%%%%%%  Theorem  %%%%%%%%%%%%%%%%%%%%%%%%%%%%%%%%%%%%%%%%%%%%%%%%%%%%%%%%%%%%%%%%%%%%%%%%%%%%%%%%%%%%%%%%%%%%%%%%%%%%%%%%%%%%%%%%%%
% \hspace*{0.2 in}The following theorem shows that s-IFRA(R) ordering is preserved under strictly increasing positive transformation. The proof could be done in the same line as it is done in Theorem \ref{thm2*2}.
% \begin{thm}
%  $X\leq_{s-IFRA(R)}Y$ if, and only if, $g(X)\leq_{s-IFRA(R)}g(Y)$ for all strictly increasing positive function $g(\cdot)$.\hfill$\Box$  
% \end{thm}
% \hspace*{0.2 in}We have the following corollary for some special types of increasing functions.
% \begin{coro}
% \begin{enumerate}
%  \item [$(a)$] $X\leq_{s-IFRA(R)}Y$ if, and only if, $pX\leq_{s-IFRA(R)}pY$ for all $p>0$.
%  \item [$(b)$] $X\leq_{s-IFRA(R)}Y$ if, and only if, $X^r\leq_{s-IFRA(R)}Y^r$ for all $r>0$.\hfill$\Box$  
%  \end{enumerate}
% \end{coro}
\hspace*{0.2 in}The following theorem shows that s-IFRA(R) ordering is location and scale invariant. The proof follows from (\ref{rre19}).
\begin{thm}
 $X\leq_{s-IFRA(R)}Y$ if, and only if, $\left(aX+b\right)\leq_{s-IFRA(R)}\left(aY+b\right)$, for any real numbers $a\;(>0)$ and $b$.\hfill $\Box$
\end{thm}
%%%%%%%%%%%%%%%%%%%%%%%%%%%%%%%%%%%%%%%%%%%%%%%%%%%%%%%%%%%%%%%%%%%%%%%%%%%%%%%%%%%%%%  s-AFQ %%%%%%%%%%%%%%%%%%%%%%%%%%%%%%%%%%%%%%%%%%%%%%%%%%
\section{s-NBU(R) Ordering}\label{sec4}
\hspace*{0.2 in}We start this section with the following definition of the s-NBU(R) ordering. 
\begin{definition} \label{def4*1} For any positive integer $s$, $X$ (or its distribution $F_X$) is said to be more $s$-NBU(R) than  
$Y$ (or its distribution $F_Y$) (written as $X\leq_{s-NBU(R)}Y$) if the random variable 
$\Phi^s_{X,Y}\;\text{has a NBU distribution}$.\hfill $\Box$
\end{definition}
\begin{remark}
For $s=1$, Definition \ref{def4*1} gives $X\leq_{su}Y$, as discussed in Sengupta and Deshpande~\cite{sd2}.\hfill $\Box$
\end{remark}
\hspace*{0.2 in}In the following proposition we give some equivalent representations of the s-NBU(R) ordering.
%%%%%%%%%%%%%%%%%%%%%%%%%%%%%%%%%%%%%%%%%%%%%%%%%%%%%%%%%%%%%%%%%%%%%%%%%%%%%%%%%%%%%%%%%%%%%%%  Proposition   %%%%%%%%%%%%%%%%%%%%%%%%%%%%%%%%%%%%%%%%%%%%%%%%%%%%%%%%%%%%%%%%%%%%%
\begin{pro}
 Definition \ref{def4*1} can equivalently be written in one of the following forms: 
 \begin{enumerate}
 \item [$(i)$] $\La_{X,s}\left(\La^{-1}_{Y,s}(x)\right)$ is super-additive  in $x\geq 0.$
 \item [$(ii)$] $\ov T^{-1}_{X,s}\left(\frac{\ov T_{X,s}(x+t)}{\ov T_{X,s}(t)}\right)\geq \ov T^{-1}_{Y,s}\left(\frac{\ov T_{Y,s}(x+t)}{\ov T_{Y,s}(t)}\right)$, for all $x,t>0.$
% \item [$(c)$] $\frac{\mu_{X,s-1}(x)}{\mu_{Y,s-1}(x)}\;\text{is decreasing in}\;x\geq0.$
 \item [$(iii)$] $\Lambda_{X,s}(Y_s)\;\text{has a NWU distribution.}$
  \end{enumerate}
\end{pro}
{\bf Proof:} The equivalence of Definition~\ref{def4*1} and $(i)$ follows from (\ref{rre14}).
Again, $(i)$ holds if, and only if, for all $a,b>0$,
$$\La_{X,s}\left(\La^{-1}_{Y,s}(a+b)\right)\geq \La_{X,s}\left(\La^{-1}_{Y,s}(a)\right)+\La_{X,s}\left(\La^{-1}_{Y,s}(b)\right),$$
or equivalently,
$$-\log\left(\frac{\ov T_{X,s}\left(\La^{-1}_{Y,s}(a+b)\right)}{\ov T_{X,s}\left(\La^{-1}_{Y,s}(a)\right)}\right)\geq -\log \ov T_{X,s}\left(\La^{-1}_{Y,s}(b)\right).$$
It can equivalently be written as
$$\ov T_{X,s}^{-1}\left(\frac{\ov T_{X,s}\left(\La^{-1}_{Y,s}(a+b)\right)}{\ov T_{X,s}\left(\La^{-1}_{Y,s}(a)\right)}\right)\geq \La^{-1}_{Y,s}(b).$$
Writing $a=\La_{Y,s}(t),\;a+b=\La_{Y,s}(x+t)$ in the above inequality, we have, for $x,t>0$,
\begin{eqnarray*}
\ov T^{-1}_{X,s}\left(\frac{\ov T_{X,s}(x+t)}{\ov T_{X,s}(t)}\right)&\geq& \La^{-1}_{Y,s}\left(-\log \frac{\ov T_{Y,s}(x+t)}{\ov T_{Y,s}(t)}\right)
\\&=&\ov T^{-1}_{Y,s}\left(\frac{\ov T_{Y,s}(x+t)}{\ov T_{Y,s}(t)}\right).
\end{eqnarray*}
Thus, the equivalence of $(i)$ and $(ii)$ is proved. The proof of $(iii)$ follows in the same line as is done in $(ii)$.\hfill$\Box$
%%%%%%%%%%%%%%%%%%%%%%%%%%%%%%%%%%%%%%%%%%%%%%%%%%%%  Lemma %%%%%%%%%%%%%%%%%%%%%%%%%%%%%%%%%%%%%%%%%%%%%%%%%%%%%%%%%%%%%%%%%%%%%%%%%%%%%%%
\\\hspace*{0.2 in} To prove the next theorem we use two lemmas which are given below without proof.
\begin{lem}\label{lem4-1}
 Let $f(\cdot)$ be a nonnegative, increasing, and super-additive function. Then $f^{-1}(\cdot)$ is sub-additive.\hfill$\Box$
\end{lem}
% {\bf Proof:} Given that $f(x+y)\geq f(x)+f(y)$. Write $f(x)=\alpha$ and $f(y)=\beta$. Then 
% \begin{eqnarray*}
%  f^{-1}(\alpha+\beta)&=&f^{-1}(f(x)+f(y)\\
%  &\leq& f^{-1}(f(x+y))\\
%  &=&x+y\\
%  &=&f^{-1}(\alpha)+f^{-1}(\beta).
%  \end{eqnarray*}\hfill$\Box$
%%%%%%%%%%%%%%%%%%%%%%%%%%%%%%%%%%%%%%%%%%%%%%%%%%%%%%%%%%%%%%%%%%%%%%%%%%%%%%%%%%%%%%%%%%%%%%%%%%%%%%%%%%%%%%%%%%%%%%%%%%%%%%%%%%%%%%%%%%%%%%%%%%%%%%%%%%
\begin{lem}\label{lem4-2}
 Let $f(\cdot)$ and $g(\cdot)$ be two nonnegative, increasing, and super-additive functions. Then $f(g(\cdot))$ is super-additive.\hfill$\Box$
\end{lem}
% {\bf Proof:} Given that 
% \begin{eqnarray}\label{geq1}
% f(x+y)\geq f(x)+f(y)
% \end{eqnarray}
% and 
% \begin{eqnarray*}\label{geq2}
% g(x+y)\geq g(x)+g(y).
% \end{eqnarray*}
% Then 
% \begin{eqnarray*}
% f(g(x+y))&\geq& f(g(x)+g(y))
% \\&\geq& fog(x)+fog(y),
% \end{eqnarray*}
% where the first inequality holds because $f$ is increasing function and the second inequality follows from (\ref{geq1}).
% Thus, $fog$ is superadditive.\hfill$\Box$
%%%%%%%%%%%%%%%%%%%%%%%%%%%%%%%%%%%%%%%%%%%%%%%%%%%%%%%%%%%%%%%%%%%%%%%%%%%%%%%%%%%%  Lemma %%%%%%%%%%%%%%%%%%%%%%%%%%%%%%%%%%%%%%%%%%%%%%%%%%%%%%%%%%%%%%%%%%%%%%%%%%%%%%%%%%%%
%%%%%%%%%%%%%%%%%%%%%%%%%%%%%%%%%%%%%%%%%%%%%%%%%%%%%%%%%%%%%%%%%%%%   Theorem   %%%%%%%%%%%%%%%%%%%%%%%%%%%%%%%%%%%%%%%%%%%%%%%%%%%%%%%%%%%%%%%%%%%%%%%%%%%%%%%%%%%%%%%%%%%%%%%%%%%%%%%%55555
\hspace*{0.2 in}The following theorem discusses some properties of the s-NBU(R) ordering.
\begin{thm}\label{th4*1}
For any positive integer $s$,
\begin{enumerate}
%\vspace*{2 in}
\item [$(i)$] $X\leq_{s-NBU(R)}X$.
 \item [$(ii)$] $X\leq_{s-NBU(R)}Y$ and $Y\leq_{s-NBU(R)}X$ hold simultaneously if, and only if, $\La_{X,s}(x)=\theta\La_{Y,s}(x)$, for some $\theta>0$ and for all $x\geq 0.$
 \item [$(iii)$] If $X\leq_{s-NBU(R)}Y$ and $Y\leq_{s-NBU(R)}Z$ then $X\leq_{s-NBU(R)}Z$.
\end{enumerate}
\end{thm}
{\bf Proof:} It is easy to verify $(i)$.
Let $X\leq_{s-NBU(R)}Y$. Then
$${\La}_{X,s}\left({\La}^{-1}_{Y,s}(x)\right)\;\text{is super-additive}.$$
By Lemma \ref{lem4-1}, the above statement can equivalently be written as 
\begin{equation}
{\La}_{Y,s}\left({\La}^{-1}_{X,s}(x)\right)\;\text{is sub-additive}.\label{eqn4-1}
\end{equation}
Further,
$Y\leq_{s-NBU(R)}X$ gives that
\begin{equation}
{\La}_{Y,s}\left({\La}^{-1}_{X,s}(x)\right)\;\text{is super-additive}.\label{eqn4-2}
\end{equation}
Combining ($\ref{eqn4-1}$) and ($\ref{eqn4-2}$), we have
$$\La_{Y,s}\left(\La^{-1}_{X,s}(x)\right)=\frac{x}{\theta},$$
for some constant $\theta\;(> 0)$.
Thus, $\La_{X,s}(x)=\theta\La_{Y,s}(x)$.
The proof of $(iii)$ follows from Lemma \ref{lem4-2}.\hfill$\Box$
%%%%%%%%%%%%%%%%%%%%%%%%%%%%%%%%%%%%%%%%%%%%%%%%%%%%%%%%%%%%    Theorem  %%%%%%%%%%%%%%%%%%%%%%%%%%%%%%%%%%%%%%%%%%%%%%%%%%%%%%%%%%%%%%%%%%
\\\hspace*{0.2 in}In the following theorem we represent the relationship between s-NBU(R) ordering and s-NBU ageing. The proof follows from Lemma~\ref{lem2-3}.
\begin{thm}
 If $\ov F_Y(x)=e^{-\la x}$, then $X\leq_{s-NBU(R)}Y$ if, and only if, $X$ is s-NBU.
\end{thm}
\hspace*{0.2 in}Since every star-shaped function is super-additive, we have the following theorem.
\begin{thm}
 If $X\leq_{s-IFRA(R)}Y$ then $X\leq_{s-NBU(R)}Y.$\hfill $\Box$
\end{thm}
%%%%%%%%%%%%%%%%%%%%%%%%%%%%%%%%%%%%%%%%%%%%%%%%%%%%%%%%%%%%%%%%%%%%  Theorem  %%%%%%%%%%%%%%%%%%%%%%%%%%%%%%%%%%%%%%%%%%%%%%%%%%%%%%%%%%%%%%%%%%%%%%%%%%%%%%%%%%%%%%%%%%%%%%%%%%%%%%%%%%%%%%%%%%
% \hspace*{0.2 in}In the following theorem we show that s-NBU(R) ordering is preserved under strictly increasing positive transformation. 
% The proof follows from (\ref{rre19}).
% \begin{thm}
%  $X\leq_{s-NBU(R)}Y$ if, and only if, $g(X)\leq_{s-NBU(R)}g(Y)$ for all strictly increasing positive function $g(\cdot)$. \hfill $\Box$
% \end{thm}
% \hspace*{0.2 in}The following corollary is a immediate consequence of the above theorem.
% \begin{coro}
% \begin{enumerate}
%  \item [$(a)$] $X\leq_{s-NBU(R)}Y$ if, and only if, $pX\leq_{s-NBU(R)}pY$ for all $p>0$.
%  \item [$(b)$] $X\leq_{s-NBU(R)}Y$ if, and only if, $X^r\leq_{s-NBU(R)}Y^r$ for all $r>0$.\hfill $\Box$
%  \end{enumerate}
% \end{coro}
\hspace*{0.2 in}The following theorem shows that s-NBU(R) ordering is scale and base invariant. The proof follows from (\ref{rre19}).
\begin{thm}
 $X\leq_{s-NBU(R)}Y$ if, and only if, $\left(aX+b\right)\leq_{s-NBU(R)}\left(aY+b\right)$, for any real numbers $a\;(>0)$ and $b$.\hfill $\Box$
\end{thm}
%%%%%%%%%%%%%%%%%%%%%%%%%%%%%%%%%%%%%%%%%%%%%%%%%%%%%%%%%%%%%%%%%%   NBUFR  %%%%%%%%%%%%%%%%%%%%%%%%%%%%%%%%%%%%%%%%%%%%%%%%%%%%%%%%%%%%%%%%%%%%%%%%%%%%%%%%%%
\section{s-NBUFR(R) Ordering}\label{sec5}
\hspace*{0.2 in}We discuss s-NBUFR(R) ordering in this section. 
\begin{definition} \label{def5*1} For any positive integer $s$, $X$ (or its distribution $F_X$) is said to be more $s$-NBUFR(R) than  
$Y$ (or its distribution $F_Y$) (written as $X\leq_{s-NBUFR(R)}Y$) if the random variable 
$\Phi^s_{X,Y}\;\text{has a NBUFR distribution}$.\hfill $\Box$
\end{definition}
%%%%%%%%%%%%%%%%%%%%%%%%%%%%%%%%%%%%%%%%%%%%%%%%%%%%%%%%%%%%%%%%%%%%%%%%%%%%%%%%%%%%%%%%%%%%%%%  Proposition   %%%%%%%%%%%%%%%%%%%%%%%%%%%%%%%%%%%%%%%%%%%%%%%%%%%%%%%%%%%%%%%%%%%%%
\hspace*{0.2 in}In the following proposition we give some equivalent conditions of the s-NBUFR(R) ordering.
\begin{pro}
 For $s=2,3\dots,$ Definition \ref{def5*1} can equivalently be written in one of the following forms:
 \begin{enumerate}
 \item [$(i)$] $\frac{r_{X,s}(x)}{r_{Y,s}(x)}\geq \frac{\tilde \mu_{Y,s-1}}{\tilde \mu_{X,s-1}}$, for all $x\geq 0$.
 \item [$(ii)$] $\frac{\mu_{Y,s-1}(x)}{\mu_{X,s-1}(x)}\geq \frac{\tilde \mu_{Y,s-1}}{\tilde \mu_{X,s-1}}$, for all $x\geq 0$.
 \item [$(iii)$] $\Lambda_{X,s}(Y_s)\;\text{has a NWUFR distribution}$.
  \end{enumerate}
\end{pro}
{\bf Proof:} $\Phi^s_{X,Y}$ is NBUFR if, and only if, for all $x\geq 0$,
$$r_{\Phi^s_{X,Y}}(x)\geq r_{\Phi^s_{X,Y}}(0),$$
or equivalently, 
\begin{eqnarray*}\label{rre9}
\frac{\ov T_{X,s-1}\left(\ov T^{-1}_{Y,s}(u)\right)}{\ov T_{X,s}\left(\ov T^{-1}_{Y,s}(u)\right)}\geq \left(\frac{\ov T_{X,s-1}(0)}{\ov T_{Y,s-1}(0)}\right)\left(\frac{\ov T_{Y,s-1}\left(\ov T^{-1}_{Y,s}(u)\right)}{u}\right),\;\text{for all}\;u\in(0,1],
\end{eqnarray*}
which holds if, and only if, 
\begin{eqnarray}\label{rre13}
 \frac{\ov T_{X,s-1}\left(x\right)}{\ov T_{X,s}\left(x\right)}\geq \frac{\ov T_{Y,s-1}\left(x\right)}{\ov T_{Y,s}(x)},\;\text{for all}\;x\geq0.
\end{eqnarray}
This can equivalently be written as
$$\frac{r_{X,s}(x)}{r_{Y,s}(x)}\geq \frac{\tilde \mu_{Y,s-1}}{\tilde \mu_{X,s-1}},\;\text{for all}\;x\geq 0. $$
Thus, the equivalence of Definition~\ref{def5*1} and $(i)$ is established.
The equivalence of $(i)$ and $(ii)$ follows from (\ref{mre5}). The proof of the equivalence of $(i)$ and $(iii)$ is obvious.\hfill $\Box$
\begin{remark}
 For $s=1$, Definition \ref{def5*1} can equivalently be written in one of the following forms:
 \begin{enumerate}
 \item [$(i)$] $\frac{r_{X,1}(x)}{r_{Y,1}(x)}\geq \frac{f_X(0)}{f_Y(0)}$, for all $x\geq 0$.
 \item [$(ii)$] $\Lambda_{X,1}(Y)\;\text{has a NWUFR distribution}$.\hfill $\Box$
  \end{enumerate}
\end{remark}
%%%%%%%%%%%%%%%%%%%%%%%%%%%%%%%%%%%%%%%%%%%%%%%%%%%%%%%%%%%%%%%%%%%%   Theorem   %%%%%%%%%%%%%%%%%%%%%%%%%%%%%%%%%%%%%%%%%%%%%%%%%%%%%%%%%%%%%%%%%%%%%%%%%%%%%%%%%%%%%%%%%%%%%%%%%%%%%%%%55555
\hspace*{0.2 in}The following theorem discusses some properties of the s-NBUFR(R) ordering.
\begin{thm}\label{th5*1}
For any positive integer $s$,
\begin{enumerate}
%\vspace*{2 in}
\item [$(i)$] $X\leq_{s-NBUFR(R)}X$.
 \item [$(ii)$] $X\leq_{s-NBUFR(R)}Y$ and $Y\leq_{s-NBUFR(R)}X$ hold simultaneously if, and only if, $\La_{X,s}(x)=\theta\La_{Y,s}(x)$, for some $\theta>0$ and for all $x\geq 0.$
 \item [$(iii)$] If $X\leq_{s-NBUFR(R)}Y$ and $Y\leq_{s-NBUFR(R)}Z$ then $X\leq_{s-NBUFR(R)}Z$.
\end{enumerate}
\end{thm}
{\bf Proof:} The proof of $(i)$ is obvious. To prove $(ii)$ we proceed as follows.
 On using (\ref{rre13}), $X\leq_{s-NBUFR(R)}Y$ reduces to the fact that, for all $x\geq0$,
\begin{eqnarray}\label{rre10}
\frac{\ov T_{X,s-1}\left(x\right)}{\ov T_{X,s}\left(x\right)}\geq \left(\frac{\ov T_{X,s-1}(0)}{\ov T_{Y,s-1}(0)}\right)\left(\frac{\ov T_{Y,s-1}\left(x\right)}{\ov T_{Y,s}(x)}\right),
\end{eqnarray}
and $Y\leq_{NBUFR(R)}X$ gives 
$$\frac{\ov T_{Y,s-1}\left(x\right)}{\ov T_{Y,s}\left(x\right)}\geq \left(\frac{\ov T_{Y,s-1}(0)}{\ov T_{X,s-1}(0)}\right)\left(\frac{\ov T_{X,s-1}\left(x\right)}{\ov T_{X,s}(x)}\right),$$
or equivalently,
\begin{eqnarray}\label{rre11}
\frac{\ov T_{X,s-1}\left(x\right)}{\ov T_{X,s}\left(x\right)}\leq \left(\frac{\ov T_{X,s-1}(0)}{\ov T_{Y,s-1}(0)}\right)\left(\frac{\ov T_{Y,s-1}\left(x\right)}{\ov T_{Y,s}(x)}\right).
\end{eqnarray}
Combining (\ref{rre10}) and (\ref{rre11}), we have
\begin{eqnarray}\label{rre12}
 \frac{\ov T_{X,s-1}\left(x\right)}{\ov T_{X,s}\left(x\right)}= \left(\frac{\ov T_{X,s-1}(0)}{\ov T_{Y,s-1}(0)}\right)\left(\frac{\ov T_{Y,s-1}\left(x\right)}{\ov T_{Y,s}(x)}\right).
\end{eqnarray}
Again, from (\ref{rre14}) and (\ref{rre15}) we have
\begin{eqnarray}
 \frac{d}{dx}\left(\La_{X,s}\La^{-1}_{Y,s}(x)\right)&=&\left(\frac{\tilde \mu_{Y,s-1}}{\tilde \mu_{X,s-1}}\right)\left(\frac{\ov T_{X,s-1}\ov T^{-1}_{Y,s}\left(e^{-x}\right)}{\ov T_{X,s}\ov T^{-1}_{Y,s}\left(e^{-x}\right)}\right)\left(\frac{e^{-x}}{\ov T_{Y,s-1}\ov T^{-1}_{Y,s}\left(e^{-x}\right)}\right)\nonumber
\\&=&\frac{1}{\theta},\label{rre16}
 \end{eqnarray}
 where  
 $$\theta=\left(\frac{\tilde \mu_{X,s-1}}{\tilde \mu_{Y,s-1}}\right)\left(\frac{\ov T_{Y,s-1}(0)}{\ov T_{X,s-1}(0)}\right),$$
 and the second equality follows from (\ref{rre12}).
 Hence, from (\ref{rre16}) we have
 $$\La_{X,s}\left(\La^{-1}_{Y,s}(x)\right)=\frac{x}{\theta},$$ or equivalently, $$\La_{X,s}(x)=\theta\La_{Y,s}(x).$$ Thus, $(ii)$ is proved.
 Again, $X\leq_{s-NBUFR(R)}Y$ gives that, for all $x\geq0$,
\begin{eqnarray}\label{rre17}
\frac{\ov T_{X,s-1}\left(x\right)}{\ov T_{X,s}\left(x\right)}\geq \left(\frac{\ov T_{X,s-1}(0)}{\ov T_{Y,s-1}(0)}\right)\left(\frac{\ov T_{Y,s-1}\left(x\right)}{\ov T_{Y,s}(x)}\right),
\end{eqnarray}
and $Y\leq_{s-NBUFR(R)}Z$ gives 
\begin{eqnarray*}
\frac{\ov T_{Y,s-1}\left(x\right)}{\ov T_{Y,s}\left(x\right)}\geq \left(\frac{\ov T_{Y,s-1}(0)}{\ov T_{Z,s-1}(0)}\right)\left(\frac{\ov T_{Z,s-1}\left(x\right)}{\ov T_{Z,s}(x)}\right),
\end{eqnarray*}
or equivalently,
\begin{eqnarray}\label{rre18}
 \left(\frac{\ov T_{X,s-1}(0)}{\ov T_{Y,s-1}(0)}\right)\left(\frac{\ov T_{Y,s-1}\left(x\right)}{\ov T_{Y,s}(x)}\right)\geq \left(\frac{\ov T_{X,s-1}(0)}{\ov T_{Z,s-1}(0)}\right)\left(\frac{\ov T_{Z,s-1}\left(x\right)}{\ov T_{Z,s}(x)}\right).
\end{eqnarray}
Thus, from (\ref{rre17}) and (\ref{rre18}) we have
\begin{eqnarray*}
\frac{\ov T_{X,s-1}\left(x\right)}{\ov T_{X,s}\left(x\right)}\geq \left(\frac{\ov T_{X,s-1}(0)}{\ov T_{Z,s-1}(0)}\right)\left(\frac{\ov T_{Z,s-1}\left(x\right)}{\ov T_{Z,s}(x)}\right).
\end{eqnarray*}
Thus, $X\leq_{s-NBUFR(R)}Z$.\hfill $\Box$
%%%%%%%%%%%%%%%%%%%%%%%%%%%%%%%%%%%%%%%%%%%%%%%%%%%%%%%%%%%%    Theorem  %%%%%%%%%%%%%%%%%%%%%%%%%%%%%%%%%%%%%%%%%%%%%%%%%%%%%%%%%%%%%%%%%%
\\\hspace*{0.2 in}In the following theorem we give a relationship between s-NBUFR(R) ordering and s-NBUFR ageing.
\begin{thm}
 If $\ov F_Y(x)=e^{-\la x}$, then $X\leq_{s-NBUFR(R)}Y$ if, and only if, $X$ is s-NBUFR.
\end{thm}\hfill $\Box$
\\\hspace*{0.2 in}Since, every NBU distribution is a NBUFR distribution, we have the following theorem.
\begin{thm}
 If $X\leq_{s-NBU(R)}Y$ then $X\leq_{s-NBUFR(R)}Y.$\hfill $\Box$
\end{thm}
% {\bf Proof:} $X\leq_{s-NBUFR(R)}Y$ gives that, for all $x,y\geq 0$,
% $$\La_{X,s}\La^{-1}_{Y,s}(x+y)-\La_{X,s}\La^{-1}_{Y,s}(x)\geq \La_{X,s}\La^{-1}_{Y,s}(y).$$
% Taking limit as $y\to 0$ on both sides of the above inequality, and then using $\La_{X,s}\La^{-1}_{Y,s}(0)=~0$, we get 
% $$\frac{d}{dx}\left(\La_{X,s}\La^{-1}_{Y,s}(x)\right)\geq \left [\frac{d}{dx}\left(\La_{X,s}\La^{-1}_{Y,s}(x)\right)\right]_{x=0},$$
% or equivalently,
% $$r_{Z^s_{X,Y}}(x)\geq r_{Z_s^*}(0).$$
% Thus, $X\leq_{s-NBUFR(R)}Y$.\hfill $\Box$
%%%%%%%%%%%%%%%%%%%%%%%%%%%%%%%%%%%%%%%%%%%%%%%%%%%%%%%%%%%%%%%%%%%%  Theorem  %%%%%%%%%%%%%%%%%%%%%%%%%%%%%%%%%%%%%%%%%%%%%%%%%%%%%%%%%%%%%%%%%%%%%%%%%%%%%%%%%%%%%%%%%%%%%%%%%%%%%%%%%%%%%%%%%%
\hspace*{0.2 in}The following theorem shows that s-NBUFR(R) ordering is scale and base invariant. 
\begin{thm}
 $X\leq_{s-NBUFR(R)}Y$ if, and only if, $\left(aX+b\right)\leq_{s-NBUFR(R)}\left(aY+b\right)$, for any real numbers $a\;(>0)$ and $b$.
\end{thm}
{\bf Proof:} For all $x\geq0$, and for any real numbers $a\;(>0)$ and $b$, the hazard rate function of the random variable $\Phi^s_{aX+b,aY+b}$ is given by
\begin{eqnarray}
r_{\Phi^s_{aX+b,aY+b}}(x)\nonumber
%\frac{d}{dx}\left(-\log \ov T_{X,s}\left(gg^{-}\ov T_{Y,s}^{-1}\left(e^{-x}\right)\right)\right)\nonum
&=&\frac{d}{dx}\left(\La_{aX+b,s}(\La_{aY+b,s})^{-1}(x)\right)\nonumber
\\&=&\frac{d}{dx}\left(\La_{X,s}\La^{-1}_{Y,s}(x)\right)\nonumber
\\&=&r_{\Phi^s_{X,Y}}(x),\label{rre20}
\end{eqnarray}
where the second equality holds from (\ref{rre19}). Thus, the result follows from Definition~\ref{def5*1}.
% \\\hspace*{0.2 in}For some specific increasing functions, the above theorem gives the following corollary.
% \begin{coro}
% \begin{enumerate}
%  \item [$(a)$] $X\leq_{s-NBUFR(R)}Y$ if, and only if, $pX\leq_{s-NBUFR(R)}pY$ for all $p>0$.
%  \item [$(b)$] $X\leq_{s-NBUFR(R)}Y$ if, and only if, $X^r\leq_{s-NBUFR(R)}Y^r$ for all $r>0$.
%  \end{enumerate}\hfill $\Box$
% \end{coro}
%%%%%%%%%%%%%%%%%%%%%%%%%%%%%%%%%%%%%%%%%%%%%%%%%%%%%%%%%%%%%%%%%%%%%%%%%%%%%%%%%%%%%%%%  NbuAFR  %%%%%%%%%%%%%%%%%%%%%%%%%%%%%%%%%%%%%%%%%%%%%%%%%%%%%%%%%%%%%%%%%%%%%%%%%%%%%%%%%%%%%%
\section{s-NBAFR(R) Ordering}\label{sec6}
\hspace*{0.2 in}We begin this section with the following definition.
\begin{definition} \label{def6*1} For any positive integer $s$, $X$ (or its distribution $F_X$) is said to be more $s$-NBAFR(R) than  
$Y$ (or its distribution $F_Y$) (written as $X\leq_{s-NBAFR(R)}Y$) if the random variable 
$\Phi^s_{X,Y}\;\text{has a NBAFR distribution}$.\hfill $\Box$
\end{definition}
%%%%%%%%%%%%%%%%%%%%%%%%%%%%%%%%%%%%%%%%%%%%%%%%%%%%%%%%%%%%%%%%%%%%%%%%%%%%%%%%%%%%%%%%%%%%%%%  Proposition   %%%%%%%%%%%%%%%%%%%%%%%%%%%%%%%%%%%%%%%%%%%%%%%%%%%%%%%%%%%%%%%%%%%%%
\hspace*{0.2 in}Some equivalent representations of the s-NBAFR(R) ordering are discussed in the following theorem.
\begin{pro}
 For $s=2,3,\dots,$ Definition \ref{def6*1} can equivalently be written in one of the following forms:  
 \begin{enumerate}
 \item [$(i)$] $\La^{-1}_{X,s}\left(x\tilde \mu_{Y,s-1} \right)\leq \La^{-1}_{Y,s}\left(x\tilde \mu_{X,s-1} \right)$, for all $x> 0$.
 \item [$(ii)$] $\frac{\La_{X,s}(x)}{\La_{Y,s}(x)}\geq \frac{\tilde \mu_{Y,s-1}}{\tilde \mu_{X,s-1}}$, for all $x> 0$.
 \item [$(iii)$] $\Lambda_{X,s}(Y_s)\;\text{has a NWAFR distribution}$.
  \end{enumerate}
\end{pro}
{\bf Proof:} $\Phi^s_{X,Y}$ has a NBAFR distribution if, and only if, for all $x>0$, 
$$-\frac{1}{x}\log \ov F_{\Phi^s_{X,Y}}(x)\geq r_{\Phi^s_{X,Y}}(0),$$
or equivalently,
\begin{eqnarray}\label{rre6}
\frac{\La_{X,s}\La^{-1}_{Y,s}(x)}{x}\geq \left(\frac{\tilde \mu_{Y,s-1}}{\tilde \mu_{X,s-1}}\right)\left(\frac{\ov T_{X,s-1}(0)}{\ov T_{Y,s-1}(0)}\right).
\end{eqnarray}
This can equivalently written as
\begin{eqnarray}\label{rre7}
\La^{-1}_{Y,s}(x)\geq \La_{X,s}^{-1}\left(x\frac{\tilde \mu_{Y,s-1}}{\tilde \mu_{X,s-1}}\right).
\end{eqnarray}
Replacing $x$ by $x\tilde \mu_{X,s-1}$ in (\ref{rre7}), we get
$$\La_{X,s}^{-1}\left(x\tilde \mu_{Y,s-1}\right)\leq \La_{Y,s}^{-1}\left(x\tilde \mu_{X,s-1}\right).$$
Thus, the equivalence of Definition~\ref{def6*1} and $(i)$ is proved. The one-to-one connection between $(i)$ and $(ii)$ follows from (\ref{rre6}).
The equivalence of $(i)$ and $(iii)$ can be proved in the same line as is done in $(i)$.\hfill $\Box$
\begin{remark}
 For $s=1,$ Definition \ref{def6*1} can equivalently be written in one of the following forms:  
 \begin{enumerate}
 \item [$(i)$] $\La^{-1}_{X,1}\left(x f_X(0) \right)\leq \La^{-1}_{Y,1}\left(x f_Y(0) \right)$, for all $x> 0$.
 \item [$(ii)$] $\frac{\La_{X,1}(x)}{\La_{Y,1}(x)}\geq \frac{f_X(0)}{\tilde f_Y(0)}$, for all $x> 0$.
 \item [$(iii)$] $\Lambda_{X,1}(Y)\;\text{has a NWAFR distribution}$.\hfill $\Box$
  \end{enumerate}
\end{remark}
%%%%%%%%%%%%%%%%%%%%%%%%%%%%%%%%%%%%%%%%%%%%%%%%%%%%%%%%%%%%%%%%%%%%   Theorem   %%%%%%%%%%%%%%%%%%%%%%%%%%%%%%%%%%%%%%%%%%%%%%%%%%%%%%%%%%%%%%%%%%%%%%%%%%%%%%%%%%%%%%%%%%%%%%%%%%%%%%%%55555
\hspace*{0.2 in}The following theorem gives some properties of the s-NBAFR(R) ordering.
\begin{thm}\label{th6*1}
For any positive integer $s$,
\begin{enumerate}
%\vspace*{2 in}
\item [$(i)$] $X\leq_{s-NBAFR(R)}X$.
 \item [$(ii)$] $X\leq_{s-NBAFR(R)}Y$ and $Y\leq_{s-NBAFR(R)}X$ hold simultaneously if, and only if, $\La_{X,s}(x)=\theta\La_{Y,s}(x)$, for some $\theta>0$ and for all $x> 0.$
 \item [$(iii)$] If $X\leq_{s-NBAFR(R)}Y$ and $Y\leq_{s-NBAFR(R)}Z$ then $X\leq_{s-NBAFR(R)}Z$.
\end{enumerate}
\end{thm}
{\bf Proof:} The proof of $(i)$ is obvious. Note that
%We only prove the result for $s=2,3,\dots.$ The result follows similarly for $s=1$. Note that
$X\leq_{s-NBAFR(R)}Y$ holds if, and only, if, for all $x>0$,
\begin{eqnarray}\label{rre1}
 {\La}_{X,s}({\La}^{-1}_{Y,s}(x))\geq \left(x\frac{\tilde \mu_{Y,s-1}}{\tilde \mu_{X,s-1}}\right)\left(\frac{\ov T_{X,s-1}(0)}{\ov T_{Y,s-1}(0)}\right).
\end{eqnarray}
Again, $Y\leq_{s-NBAFR(R)}X$ holds if, and only if, for all $x>0$,
\begin{eqnarray}\label{rre2}
 {\La}_{Y,s}({\La}^{-1}_{X,s}(x))\geq \left(x\frac{\tilde \mu_{X,s-1}}{\tilde \mu_{Y,s-1}}\right)\left(\frac{\ov T_{Y,s-1}(0)}{\ov T_{X,s-1}(0)}\right).
\end{eqnarray}
Replacing $x$ by ${\La}_{X,s}({\La}^{-1}_{Y,s}(x))$ in (\ref{rre2}), we have
%$$x\geq \left(\frac{\tilde \mu_{X,s-1}}{\tilde \mu_{Y,s-1}}\right)\left(\frac{\ov T_{Y,s-1}(0)}{\ov T_{X,s-1}(0)}\right){\La}_{X,s}(\La^{-1}_{Y,s}(x)),$$
%or equivalently,
\begin{eqnarray}\label{rre3}
 \La_{X,s}(\La^{-1}_{Y,s}(x))\leq \left(x\frac{\tilde \mu_{Y,s-1}}{\tilde \mu_{X,s-1}}\right)\left(\frac{\ov T_{X,s-1}(0)}{\ov T_{Y,s-1}(0)}\right).
\end{eqnarray}
Combining (\ref{rre1}) and (\ref{rre3}), we have
\begin{eqnarray*}
 \La_{X,s}(\La^{-1}_{Y,s}(x))=\theta x ,
\end{eqnarray*}
where $$\theta=\left(\frac{\tilde \mu_{Y,s-1}}{\tilde \mu_{X,s-1}}\right)\left(\frac{\ov T_{X,s-1}(0)}{\ov T_{Y,s-1}(0)}\right).$$ Thus, $\La_{X,s}(x)=\theta\La_{Y,s}( x)$, and hence 
$(ii)$ is proved. Again,
%Here, we also prove the result for $s=2,3,\dots$ and the result follows similarly for $s=1$.
$X\leq_{s-NBAFR}Y$ gives
\begin{equation}\label{rre4}
{\La}_{X,s}({\La}^{-1}_{Y,s}(x))\geq \left(x\frac{\tilde \mu_{Y,s-1}}{\tilde \mu_{X,s-1}}\right)\left(\frac{\ov T_{X,s-1}(0)}{\ov T_{Y,s-1}(0)}\right),
\end{equation}
and $Y\leq_{s-NBAFR}Z$ gives
\begin{equation}\label{rre5}
{\La}_{Y,s}({\La}^{-1}_{Z,s}(x))\geq \left(x\frac{\tilde \mu_{Z,s-1}}{\tilde \mu_{Y,s-1}}\right)\left(\frac{\ov T_{Y,s-1}(0)}{\ov T_{Z,s-1}(0)}\right).
\end{equation}
Now, 
\begin{eqnarray*}
 \La_{X,s}(\La^{-1}_{Z,s}(x))&=&\La_{X,s}\La^{-1}_{Y,s}\left(\La_{Y,s}\La^{-1}_{Z,s}(x)\right)
 \\&\geq&\La_{X,s}\La^{-1}_{Y,s}\left(x\frac{\tilde \mu_{Z,s-1}}{\tilde \mu_{Y,s-1}}\frac{\ov T_{Y,s-1}(0)}{\ov T_{Z,s-1}(0)}\right)
 \\&\geq&\left(x\frac{\tilde \mu_{Z,s-1}}{\tilde \mu_{Y,s-1}}\right)\left(\frac{\ov T_{Y,s-1}(0)}{\ov T_{Z,s-1}(0)}\right)\left(\frac{\tilde \mu_{Y,s-1}}{\tilde \mu_{X,s-1}}\right)\left(\frac{\ov T_{X,s-1}(0)}{\ov T_{Y,s-1}(0)}\right)
 \\&=&\left(x\frac{\tilde \mu_{Z,s-1}}{\tilde \mu_{X,s-1}}\right)\left(\frac{\ov T_{X,s-1}(0)}{\ov T_{Z,s-1}(0)}\right),
\end{eqnarray*}
where the first inequality holds from (\ref{rre5}) and using the fact that $\La_{X,s}\La^{-1}_{Y,s}(\cdot)$ is an increasing function. The second inequality
follows from (\ref{rre4}). Thus, $X\leq_{s-NBAFR(R)}Z$.\hfill $\Box$
%%%%%%%%%%%%%%%%%%%%%%%%%%%%%%%%%%%%%%%%%%%%%%%%%%%%%%%%%%%%    Theorem  %%%%%%%%%%%%%%%%%%%%%%%%%%%%%%%%%%%%%%%%%%%%%%%%%%%%%%%%%%%%%%%%%%
\\\hspace*{0.2 in} The following theorem shows that $X$ is smaller than exponential random variable in s-NBAFR(R) ordering if, and only if, $X$ has a NBAFR distribution.
The proof follows from Lemma \ref{lem2-3}.
\begin{thm}
 If $\ov F_Y(x)=e^{-\la x}$, then $X\leq_{s-NBAFR(R)}Y$ if, and only if, $X$ is s-NBAFR.
\end{thm}\hfill $\Box$
\\\hspace*{0.2 in}Since, every NBUFR distribution is a NBAFR distribution, we have the following theorem.
\begin{thm}
 If $X\leq_{s-NBUFR(R)}Y$ then $X\leq_{s-NBAFR(R)}Y.$\hfill $\Box$
\end{thm}
%%%%%%%%%%%%%%%%%%%%%%%%%%%%%%%%%%%%%%%%%%%%%%%%%%%%%%%%%%%%%%%%%%%%  Theorem  %%%%%%%%%%%%%%%%%%%%%%%%%%%%%%%%%%%%%%%%%%%%%%%%%%%%%%%%%%%%%%%%%%%%%%%%%%%%%%%%%%%%%%%%%%%%%%%%%%%%%%%%%%%%%%%%%%
% \hspace*{0.2 in}The following theorem shows that s-NBAFR ordering is preserved under strictly increasing positive transformation.
% On using (\ref{rre20}), the proof follows from Definition~\ref{def6*1}.
% \begin{thm}
%  $X\leq_{s-NBAFR(R)}Y$ if, and only if, $g(X)\leq_{s-NBAFR(R)}g(Y)$ for all strictly increasing positive function $g(\cdot)$.  \hfill $\Box$
% \end{thm} 
% \hspace*{0.2 in}The following corollary is a immediate consequence of the above theorem.
% \begin{coro}
% \begin{enumerate}
%  \item [$(a)$] $X\leq_{s-NBAFR(R)}Y$ if, and only if, $pX\leq_{s-NBAFR(R)}pY$ for all $p>0$.
%  \item [$(b)$] $X\leq_{s-NBAFR(R)}Y$ if, and only if, $X^r\leq_{s-NBAFR(R)}Y^r$ for all $r>0$.\hfill $\Box$
%  \end{enumerate}
% \end{coro}
\hspace*{0.2 in}That the s-NBAFR(R) ordering is scale and base invariant, is shown in the following theorem. On using (\ref{rre20}), the proof follows from Definition~\ref{def6*1}.
\begin{thm}
 $X\leq_{s-NBAFR(R)}Y$ if, and only if, $\left(aX+b\right)\leq_{s-NBAFR(R)}\left(aY+b\right)$, for any real numbers $a\;(>0)$ and $b$.\hfill $\Box$
\end{thm}
%%%%%%%%%%%%%%%%%%%%%%%%%%%%%%%%%%%%%%%%%%%%%%%%%%%%%%%%%%%%%%%%%%%%%%%%%%%%%%%%%%%%%%%%%%%%%%%%%%%%%%%%%%%%%%%%%%%%%%%%%%%%%%%%%%%%%%%%%%%%%%%%%%%%%%%%%%%%%%%%%%%%%%%%%%%%%%%%%%%%%%%%%%%%
                                                    %%%%%%%%%%%%%%%%%%%%%%%%%%%%%%  Conclusion %%%%%%%%%%%%%%%%%%%%%%%%%%%%%%%%%%%%%%%%
%%%%%%%%%%%%%%%%%%%%%%%%%%%%%%%%%%%%%%%%%%%%%%%%%%%%%%%%%%%%%%%%%%%%%%%%%%%%%%%%%%%%%%%%%%%%%%%%%%%%%%%%%%%%%%%%%%%%%%%%%%%%%%%%%%%%%%%%%%%%%%%%%%%%%%%%%%%%%%%%%%%%%%%%%%%%%%%%%%%%%%%%%%%55555
\section{Concluding Remarks}\label{sec7-1}
\hspace*{0.2 in}In this paper we give some new generalized stochastic orderings, and study some of their properties. These orderings may be helpful to visualize 
the positive ageing classes from different aspects.
To handle the crossing hazard rates and/or crossing mean residual lives problem, we may find out a new direction with the help of these orderings. 
This unified study is meaningful because it gives a complete scenario of the existing results (available in the literature) together with the new results.
The usefulness of relative ageing is well explained in Sengupta and Deshpande~\cite{sd2}, and Kalashnikov and Rachev~\cite{kr6}. Keeping the importance of the relative ageing in mind, 
we have taken an attempt to discuss different kinds of relative ageing in a unified way so that different kinds of relative ageing properties come under a single umbrella.
Further, the different characterizations of these relative ageing properties are important because of their theoretical insight in one hand, and the systems belonging to this 
ageing classes help the practitioners (viz. reliability and design engineers) manipulate it for its nice mathematical properties on the other. To make the usefulness of these kind of orderings
more appealing, let us take a particular example as discussed below.
\begin{example}\label{cse61}
%  \begin{figure}
% \centering
% \includegraphics[width=10.5 cm,keepaspectratio]{fig62.eps} \vspace{.5cm}
% \caption{Plot of $r_{X,1}(t)$ and $r_{Y,1}(t)$ against $t\in [0,0.7]$
% (Example $\ref{cse61}$).} \label{cxe62}
% \end{figure}
%  \begin{figure}
% \centering
% \includegraphics[width=10.5 cm,keepaspectratio]{fig61.eps} \vspace{.5cm}
% \caption{Plot of $r_{X,1}(t)/r_{Y,1}(t)$ against $t\in [0,0.4]$
% (Example $\ref{cse61}$).} \label{cxe61}
% \end{figure}
 Let $X$ be random variable having $\mu_{X,1}(t)=1/(4+11t^2),$ $t\geq 0$, and $Y$ be another random variable 
  having $\mu_{Y,1}(t)=1/(4+5t^2),$ $t\geq 0$. Then
  $$r_{X,1}(t)=4+11t^2-\frac{22t}{4+11t^2},$$
  and $$r_{Y,1}(t)=4+5t^2-\frac{10t}{4+5t^2}.$$
 By drawing the figures of $r_{X,1}(t)$ and $r_{Y,1}(t)$, it can be shown that $X$ and $Y$ have crossing hazard rates. Note that
 $$\frac{r_{X,1}(t)}{r_{Y,1}(t)}=\left(\frac{(4+11t^2)^2-22t}{(4+5t^2)^2-10t}\right)\left(\frac{4+5t^2}{4+11t^2}\right),$$
which can be shown to non-monotone, and hence $X\nleq_{1-IFR(R)}Y$.
Further $$\frac{r_{X,2}(t)}{r_{Y,2}(t)}=\frac{4+11t^2}{4+5t^2}\;\text{is increasing in}\;t,$$
and hence $X\leq_{2-IFR(R)}Y$ follows from Proposition~\ref{pp61}.$\hfill\Box$
\end{example}
In the above example we see that hazard rates of $X$ and $Y$ have crossed each other. 
So, none of the two dominates the other in terms of their failure rates.
In order to decide on the better system, i.e., to see which one is ageing slower, we take $s=1$, i.e., we compare them in terms of $1$-IFR(R) order. 
It is noted that none of the two dominates the other as far as $1$-IFR(R) order is concerned.
This means that if we concentrate our study based on $1$-IFR(R) order only, we cannot conclude which of the two is better. To overcome this difficulty we take $s=2$, which gives 
a comparison, known as $2$-IFR(R) order. Here we see that the ratio of $r_{X,2}(t)$ and $r_{Y,2}(t)$ is monotone $-$ clearly showing the dominance of one over the other.
Thus, we are now in a position to say that $X$ is ageing faster compared to $Y$. So the system with life distribution $Y$ is better. 
The above example gives the importance of $s$-IFR(R) ($s\geq 2$) order. In a similar spirit, the other generalized orders
are defined and studied to help the reliability practitioners to decide on how to choose the better one.
We conclude our discussion by mentioning the following chain
of implications of the generalized stochastic orderings.
\\\hspace*{1 in}$X\leq_{s-IFR(R)}Y\Rightarrow ~X\leq_{s-IFRA(R)}Y$
\\\hspace*{2.6 in}$\Downarrow$
\\\hspace*{2.4 in}$ X\leq_{s-NBU(R)}Y$
\\\hspace*{2.6 in}$\Downarrow$
\\\hspace*{2.4 in}$X\leq_{s-NBUFR(R)}Y\Rightarrow X\leq_{s-NBAFR(R)}Y.$
%%%%%%%%%%%%%%%%%%%%%%%%%%%%%%%%%%%%%%%%%%%%%%%%%%%%%%%%%%  Acknowledgements  %%%%%%%%%%%%%%%%%%%%%%%%%%%%%%%%%%%%%%
\section*{Acknowledgements}
%\hspace*{0.3 in}The authors gratefully acknowledge the constructive comments of the referee which lead to an improved version of the manuscript.
  \hspace*{0.3 in}The authors are thankful to the Editor, and anonymous Reviewers for their valuable constructive comments and suggestions which lead to an improved version of the manuscript.
  Financial support from Council of Scientific and Industrial Research, New Delhi (CSIR Grant No. 09/921(0060)2011-EMR-I) is sincerely
 acknowledged by Nil Kamal Hazra. 
%%%%%%%%%%%%%%%%%%%%%%%%%%%%%%%%%%%%%%%%%%%%%%%%%%%%%%%%%%%%%%%%%%%%%%%%%%%%%%%%%%%%%%%%%%%%%%%%%%%%%%%%%%%%%%%%%%%%%%%%%%%%%%%%%%%%%%%%%%%%%%%%%%%%%%%%%%%%%%%%%%%%%%%%%%%%%%%%%%%%%%555555555555555555
                                                              %%%%%%%%%%%%%%%%%%   References  %%%%%%%%%%%%%%%%%%%%%%%%%%%
%%%%%%%%%%%%%%%%%%%%%%%%%%%%%%%%%%%%%%%%%%%%%%%%%%%%%%%%%%%%%%%%%%%%%%%%%%%%%%%%%%%%%%%%%%%%%%%%%%%%%%%%%%%%%%%%%%%%%%%%%%%%%%%%%%%%%%%%%%%%%%%%%%%%%%%%%%%%%%%%%%%%%%%%%%%%%%%%%%%%%%%%%%%%%%                                                              

\end{document}